\documentclass[reprint,9pt,amsmath,amssymb,aps,superscriptaddress,longbibliography]{revtex4-2}
\usepackage{hyperref}
\usepackage{extarrows}
\usepackage{graphicx}
\usepackage{dcolumn}
\usepackage{bm}
\usepackage[dvipsnames]{xcolor}
\usepackage[]{natbib}
\usepackage{url}

\usepackage{soul}

\mathchardef\mhyphen="2D

\begin{document}
	\title{A simple model of global cascades on random hypergraphs}

    \author{Lei Chen} 
    \affiliation{Alibaba Research Center for Complexity Sciences, Hangzhou Normal University, Hangzhou, Zhejiang 311121, China}

    \author{Yanpeng Zhu}
    \affiliation{Alibaba Research Center for Complexity Sciences, Hangzhou Normal University, Hangzhou, Zhejiang 311121, China}

    \author{Jiadong Zhu}
    \affiliation{Alibaba Research Center for Complexity Sciences, Hangzhou Normal University, Hangzhou, Zhejiang 311121, China}

    \author{Zhongyuan Ruan}
    \affiliation{Institute of Cyberspace Security, Zhejiang University of Technology, Hangzhou, 310023, China}
    \affiliation{College of Information Engineering, Zhejiang University of Technology, Hangzhou 310023, China}

    \author{Michael Small}
    \affiliation{The University of Western Australia, Crawley, WA 6009, Australia}
    \affiliation{Mineral Resources, CSIRO, Kensington, WA 6151, Australia}

    \author{Kim Christensen}
    \affiliation{Blackett Laboratory and Center for Complexity Science, Imperial College London, SW7 2AZ, London, United Kingdom}
    \author{Run-Ran Liu} \email{runranliu@163.com}
    \affiliation{Alibaba Research Center for Complexity Sciences, Hangzhou Normal University, Hangzhou, Zhejiang 311121, China}

    \author{Fanyuan Meng}\email{fanyuan.meng@hotmail.com}
    \affiliation{Alibaba Research Center for Complexity Sciences, Hangzhou Normal University, Hangzhou, Zhejiang 311121, China}

\begin{abstract}
This study introduces a comprehensive framework that situates information cascades within the domain of higher-order interactions, utilizing a double-threshold hypergraph model. We propose that individuals (nodes) gain awareness of information through each communication channel (hyperedge) once the number of information adopters surpasses a threshold $\phi_m$. However, actual adoption of the information only occurs when the cumulative influence across all communication channels exceeds a second threshold, $\phi_k$. We analytically derive the cascade condition for both the case of a single seed node using percolation methods and the case of any seed size employing mean-field approximation. Our findings underscore that when considering the fractional seed size, $r_0 \in (0,1]$, the connectivity pattern of the random hypergraph, characterized by the hyperdegree, $k$, and cardinality, $m$, distributions, exerts an asymmetric impact on the global cascade boundary. This asymmetry manifests in the observed differences in the boundaries of the global cascade within the $(\phi_m, \langle m \rangle)$ and $(\phi_k, \langle k \rangle)$ planes. However, as $r_0 \to 0$, this asymmetric effect gradually diminishes.  Overall, by elucidating the mechanisms driving information cascades within a broader context of higher-order interactions, our research contributes to theoretical advancements in complex systems theory.
\end{abstract}

\maketitle
\section{Introduction}
In the intricate dynamics of human interactions, information \cite{iribarren2009impact,jiang2014evolutionary}, opinons \cite{moussaid2013social,watts2007influentials}, innovations \cite{wejnert2002integrating,karsai2014,shih2010facebook}, rumors
\cite{moreno2004dynamics,draief2010epidemics,zhao2012sihr,wang2014siraru}, and even emotions \cite{stieglitz2013emotions,espinosa2014characterizing,prollochs2021emotions} can spread rapidly, a phenomenon known as social contagion, which profoundly shapes individual and collective behaviors. In today's hyperconnected world, with its numerous social media platforms and online communities \cite{shih2010facebook,karsai2014,centola2010spread}, the potential for social contagion to exert significant influence on our lives has been amplified. Consequently, understanding the mechanisms governing this pervasive phenomenon is of paramount importance.

In recent decades, the field of network science has witnessed significant expansion, leading to the development of numerous models aimed at understanding the intricate dynamics of social contagion \cite{christakis2013social,cozzo2013contact,ugander2012structural,sharpanskykh2014modelling,castellano2009statistical}. Among these models, threshold models have emerged as a prominent paradigm \cite{granovetter1983threshold,watts2002simple,gleeson2007seed,ruan2015kinetics}. Rooted in the theory of social influence \cite{mcdonald2015social,cialdini2004social}, these models propose that individuals possess a threshold for adopting new information or behaviors, necessitating a critical proportion or number of their peers to have already adopted before they are inclined to do so themselves. Watts \cite{watts2002simple} introduced a threshold model to explore the effect of network connectivity on information cascades within simple networks, while Gleeson and Cahalane
 \cite{gleeson2007seed} demonstrated that the size of the initial seed in this model influences the final cascade size significantly.

However, modeling social contagion solely through simple pairwise interactions may obscure the influence of group dynamics, thereby limiting our understanding of the phenomenon. Hypergraph models offer a natural way to represent these group or higher-order interactions and capture the intricate dependencies among individuals. For example, Liu \textit{et al.} \cite{liu2023threshold} demonstrated how the collapse of a group can be triggered by the failure of its members to reach a threshold, leading to a cascading failure phenomenon. Additionally, Arruda \textit{et al.} \cite{de2020social,ferraz2023multistability} proposed models where agents transition into contagion states based on probabilities when the number of contagion agents exceeds certain thresholds.

Nonetheless, individuals are often subject to concurrent influences from multiple social groups, each with its own distinct norms or pressures. For instance, individuals may concurrently engage with familial, professional, and interest-based communities. Xu \textit{et al.} \cite{xu2022dynamics} proposed a model wherein a node activates only if the fraction of active neighboring nodes across all groups exceeds a certain threshold. However, this approach overlooks the multi-stage nature of social influence on individual information adoption. Specifically, there is the dissemination and internalization of new ideas, behaviors, or emotions within the group. This is followed by the individual's eventual adoption or rejection of the proposed behavior under the influence of all group norms or pressures collectively and concurrently.

We introduce a comprehensive framework that contextualizes information cascades within the domain of higher-order interactions, employing a double-threshold hypergraph model. In this framework, we assume that individuals (nodes) become aware of information through each communication channel (hyperedge) once the number of adopters exceeds a threshold $\phi_m$. However, the actual adoption of a piece of information occurs only when the cumulative influence across all communication channels exceeds a second threshold, $\phi_k$. 

Our framework facilitates the analytical derivation of the information cascade condition, addressing both the case of a single seed node using percolation methods and the case of any seed size through mean-field approximation. Additionally, our study elucidates the intricate interplay between the initial seed size, the network structure, and the dynamics of information cascade. Through this research, we endeavor to enhance our understanding of the mechanisms underlying social contagion cascade processes with higher-order interactions.

\section{Model}
We introduce a random double-threshold hypergraph model to investigate how the higher-order interactions affect an information cascade process. The random hypergraph is characterized by a set of $N$ nodes $V = \{v_i\}$ where $i \in \{1,2,\dots,N\}$, and a collection of $M$ hyperedges $E = \{e_j\}$ with $j \in \{1,2,\dots,M\}$. The hyperdegree of each node refers to the number of hyperedges incident to that node, following the Poisson distribution $p_k = \frac{e^{\langle k \rangle}{\langle k \rangle}^k}{k!}$ with average hyperdegree $\langle k \rangle$. The cardinality of each hyperedge refers to the number of nodes that belong to it, following the Poisson distribution $q_m = \frac{e^{\langle m \rangle}{\langle m \rangle}^m}{m!}$ with average cardinality $\langle m \rangle$. 


We assume that nodes and hyperedges in the model can exist in two states: 0 for inactive and 1 for active. For any node $v_i$, the active state is indicated by $s_i=1$, and the inactive state by $s_i=0$. Additionally, for a hyperedge $e_j$, the active state is $h_j=1$ and the inactive state is $h_j=0$. We define a variable $y_i=\sum\limits_{j, {v_i \in e_j}} h_j$ for each node $v_i$, representing the number of active hyperedges surrounding the node. Similarly, for each hyperedge $e_j$, the variable $x_j=\sum\limits_{i, {v_i \in e_j}} s_i$ indicates the number of active nodes it contains. 

\begin{figure*}
    \centering
    \includegraphics[width=\linewidth]{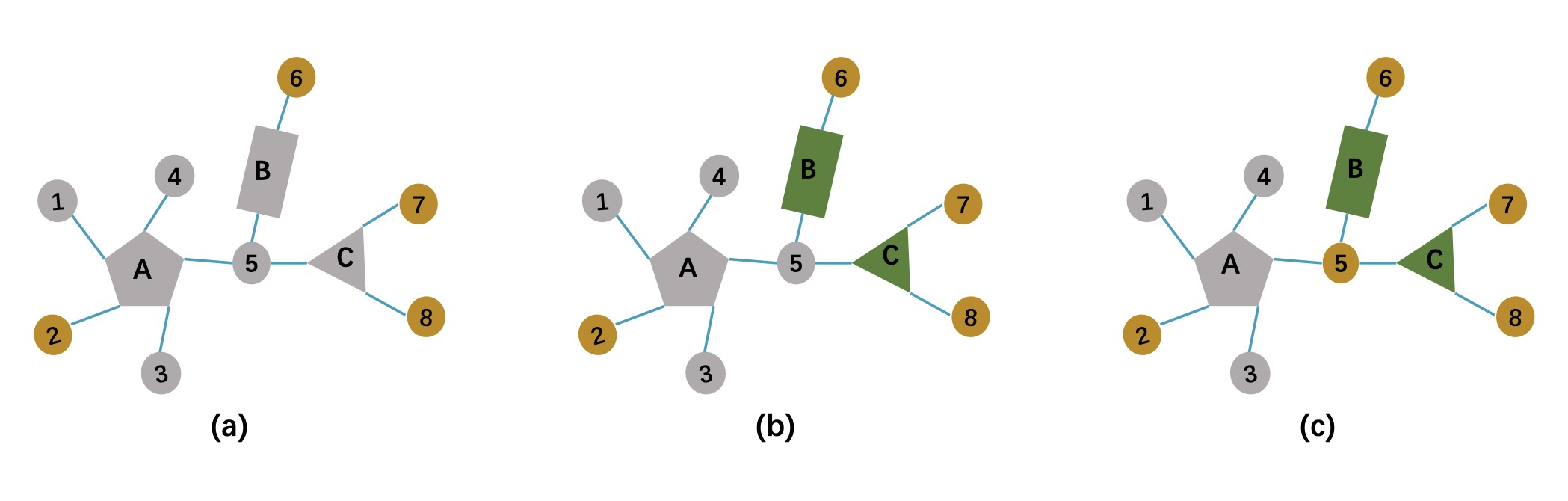}
    \caption{A visual representation of the information cascade process within a random hypergraph with double thresholds $\phi_k$ and $\phi_m$. In this illustration, circles represent nodes (individuals) numbered 1 to 8, while polygons denote hyperedges (groups) identified as $A$ ($m=5$), $B$ ($m=2$), and $C$ ($m=3$). Inactive nodes and hyperedges are displayed in gray, whereas active nodes are highlighted in brown, and active hyperedges are portrayed in green. The schematic showcases the following cascade process: Initially, nodes 2, 6, 7, and 8 serve as seed nodes in stage (a). With a threshold $\phi_m=0.45$, hyperedges $B$ ($\frac{x_B}{m}=\frac{1}{2}\geq \phi_m$) and $C$ ($\frac{x_C}{m}=\frac{2}{3}\geq \phi_m$) become activated while hyperedge $A$ ($\frac{x_A}{m}=\frac{1}{5}< \phi_m$) remains inactive in stage (b). Subsequently, with $\phi_k$ at 0.5, node 5 ($\frac{y_5}{k}=\frac{2}{3}\geq \phi_k$) transitions to an active state in stage (c). The fraction of active nodes for hyperedge $A$ is now $2/5 < \phi_m$, so the cascade comes to an end.}
    \label{fig:illustrate}
\end{figure*}

Our framework identifies individuals as nodes and groups as hyperedges to simulate information cascade dynamics. The process begins with a small fraction $r_0$ of nodes in the active state, serving as the seed. The state transition mechanism is twofold: 1) 
A hyperedge becomes active from an inactive state when the fraction of its active nodes (individuals) exceeds a threshold $0 \leq \phi_m \leq 1$, i.e., $x_j/m \geq \phi_m $, representing the minimal influence necessary for an individual to become aware of information through each communication channel; 2) An inactive node becomes active if the fraction of active hyperedges (groups) it connects to surpasses another threshold $0 \leq \phi_k \leq 1$, i.e., $y_i/k \geq \phi_k$, signifying the cumulative influence needed for each node's activation through collective communication channels. The information cascade progresses until the system reaches a stable state, where no additional nodes or hyperedges change to an active state (see an example in Fig.~\ref{fig:illustrate}).

\section{Results} 
\subsection{Single Seed Node}
To explore the effect of a minimal initial disturbance, characterized by a single node, on the information cascade processes, we set the initial fractional size as $r_0 = 1/N \to 0$.

Building upon the insights of  \cite{watts2002simple}, We define a node with hyperdegree $k$ as \emph{vulnerable} when the probability $\rho_k = P\left(1/k \geq \phi_k\right)$ is satisfied. This implies that a single active hyperedge can activate the node. The complementary state is termed \emph{stable}. Consequently, the probability of a specific vulnerable node $v_i$ possessing a hyperdegree of $k$ is represented by $\rho_k p_k$. The generating function for vulnerable nodes is
\begin{equation}
    G_{k0}(x)=\sum_{k=0}\rho_k p_{k} x^{k},
\end{equation}
where 
\begin{equation}
\rho_k=\left\{
     \begin{array}{ll}
     0, & \text{ if } k>0 \text{ and } 1/k < \phi_k , \\
     
     1, & \text{ if } k>0 \text{ and } 1/k \geq \phi_k, \text{ or } k = 0.\\
     \end{array}
\right.
\end{equation}

Moreover, the generating function for a vulnerable node reached by a random hyperedge can be given by
\begin{equation}
    G_{k1}(x)=\sum_{k=0} \frac{k \rho_k p_{k}}{\langle k \rangle} x^{k-1} =  \frac{G_{k0}^{\prime}(x)}{\langle k \rangle}.
\end{equation}

Similarly, we define a hyperedge with cardinality $m$ as \emph{vulnerable} when the probability $\beta_m = P(1/m \geq \phi_m)$ is satisfied. This indicates that a single active node can activate the hyperedge. The complementary state is termed \emph{stable}. Consequently, the probability of a hyperedge being both vulnerable and possessing a cardinality of $m$ is given by $\beta_m q_m$. The generating function for vulnerable hyperedges is 
\begin{equation}
    G_{m0}(x)=\sum_{m=0}\beta_m q_{m} x^{m},
\end{equation}
where
\begin{equation}
\beta_m=\left\{
     \begin{array}{ll}
     0, & \text{ if } m>0 \text{ and } 1/m < \phi_m , \\
     
     1, &  \text{ if } m>0 \text{ and } 1/m \geq \phi_m, \text{ or } m=0.\\
     \end{array}
\right.
\end{equation}

Furthermore, as the cardinality of a hyperedge increases, the probability of a randomly chosen node being a part of it also increases. The probability of a random node being a member of a hyperedge is directly proportional to $mq_m$. Therefore, the generating function for a vulnerable hyperedge reached by a random associated node can be given by
\begin{equation}
    G_{m1}(x)= \sum_{m=1} \frac{m \beta_m q_{m}}{\langle m \rangle } x^{m-1} = \frac{G_{m0}^{\prime}(x)}{\langle m \rangle }.
    \label{eq:G_m1}
\end{equation}

In our model, \emph{the vulnerable cluster} can be delineated by the ensemble of vulnerable nodes and hyperedges, thereby implying that an information cascade can be initiated by a single seed node within the cluster. 

We first introduce the generating function
\begin{equation}
    H_{1}(x)=\sum_{n=0}\eta_n x^{n},
\end{equation}
where $\eta_n$ signifies the probability that a random vulnerable hyperedge can reach a vulnerable cluster of size $n$. 

The self-consistency equation for $H_1(x)$ is derived as
    \begin{equation}
       H_{1}(x)=1-G_{k1}(1)+x G_{k1}\left(1-G_{m1}(1)+G_{m1}(H_{1}(x))\right).
       \label{eq:H_1(x)}
\end{equation}

The above equation can be interpreted as follows: The first term, $1-G_{k1}(1)$, represents the probability that a randomly chosen vulnerable hyperedge reaches a stable node, essentially indicating when the vulnerable cluster size $n=0$. The second term considers the probability that a vulnerable hyperedge reaches a vulnerable size $n \geq 1$, given that the hyperedge has already reached a vulnerable node. This probability is expressed as $xG_{k1}\left(1-G_{m1}(1)+G_{m1}(H_{1}(x))\right)$. Breaking this down further: $1-G_{m1}(1)$ signifies the probability of reaching a stable hyperedge, while $G_{m1}(H_1(x))$ represents the probability of reaching a vulnerable cluster of size $n$ conditional on reaching a vulnerable hyperedge.

To determine the size of vulnerable clusters, we introduce another generating function 
\begin{equation}
    H_{0}(x)=\sum_{n}\alpha_n x^{n},
\end{equation}
where $\alpha_n$ represents the probability that a randomly chosen node is part of a vulnerable cluster of size $n$. 

According to Eq.~\eqref{eq:H_1(x)}, $H_0(x)$ can be expressed in alternate form
\begin{equation}
    H_{0}(x)=1-G_{k0}(1)+x G_{k0}\left(1-G_{m1}(1)+G_{m1}(H_{1}(x))\right).
    \label{eq:h_0(x)}
\end{equation}
    
This equation provides insights into the distribution of vulnerable cluster size within the hypergraph. Moreover, with the help of $H_{0}(x)$, we can compute the average vulnerable cluster size $\langle n \rangle$ as follows 
\begin{equation}
\langle n \rangle = H_0^\prime (1) =  G_{k0}(1)+ G_{k0}^{\prime}(1)G_{m1}^{\prime}(1)H_{1}^{\prime}(1).
\label{eq:average_cluster}
\end{equation}

Furthermore, by utilizing Eq.~\eqref{eq:H_1(x)}, we can obtain
\begin{equation}
H_{1}^{\prime}(1)  = \frac{G_{k1}(1)}{1- G_{k1}^{\prime}(1)G_{m1}^{\prime}(1)}.
\label{eq:H_1_d}
\end{equation}

Hence, the average vulnerable cluster size $\langle n \rangle$ in Eq.~\eqref{eq:average_cluster} will diverge when the denominator in Eq.~\eqref{eq:H_1_d} $1- G_{k1}^{\prime}(1)G_{m1}^{\prime}(1)=0$.

So the \emph{cascade condition} can be given as
    \begin{equation}
    G_{k1}^{\prime}(1)G_{m1}^{\prime}(1) = \frac{\sum\limits_{k} k(k-1) \rho_{k} p_{k}}{\langle k \rangle}
\frac{\sum\limits_{m} m(m-1) \beta_{m} q_{m}}{\langle m \rangle} = 1.
    \label{eq:threshold_condition}
\end{equation}

\begin{figure*}
    \centering
    \includegraphics[width=\linewidth]{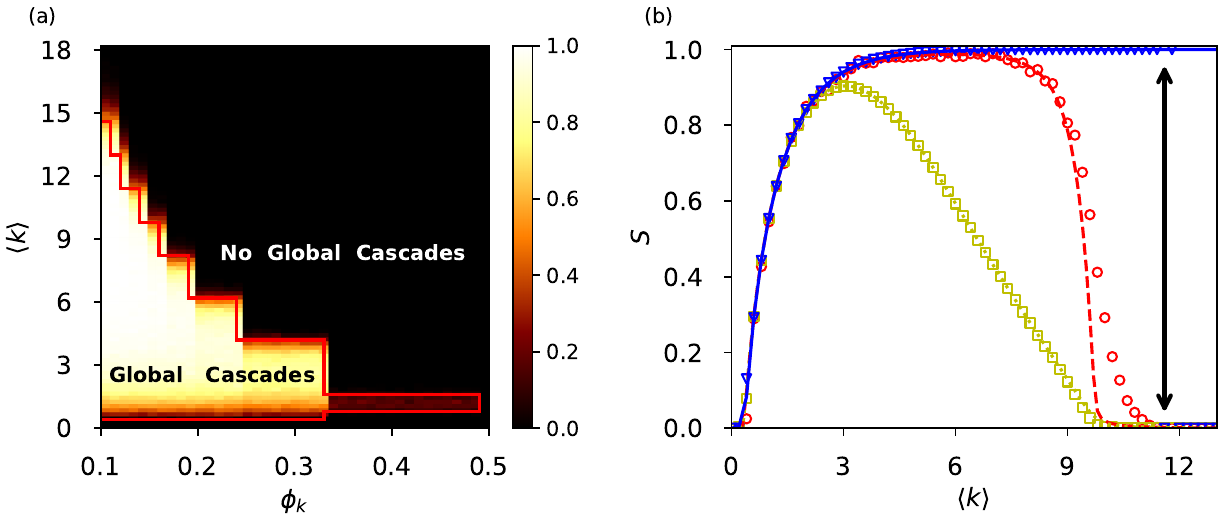}
    \caption{(a) Cascade region for the case of a single seed node in random hypergraphs. The red solid line delineates the region of the $(\phi_k, \langle k \rangle)$ plane where the cascade condition of Eq.~\eqref{eq:threshold_condition} is satisfied. The color-coded values depict simulation results indicating the average fractional cascade size exceeding 0.1 for 1000 hypergraph realizations, with parameters set to $N=10^4$, $\phi_m = 0.1$, and $\langle m \rangle = 3$; (b) Relationship between the fractional size of the giant vulnerable cluster $S_v$, the extended giant vulnerable cluster $S_e$, the giant cluster $S$, and the average hyperdegree $\langle k \rangle$. The yellow dotted line represents the exact solution derived from Eq.~\eqref{eq:S_v} in Appendix \ref{sec:app_A}, while the corresponding squares denote the average fractional size $S_v$ of the giant vulnerable cluster over 1000 hypergraph realizations. Additionally, the red dashed line demonstrates the fractional size $S_e$ of the extended giant vulnerable cluster derived from Eq.~\eqref{eq:S_e} in Appendix \ref{sec:app_A}, while the corresponding circles denote the frequency of global cascades (i.e., the ratio of 1000 hypergraph realizations in which the fractional cascade size exceeds 0.1), and the corresponding blue triangles represent the average value among the fractional cascade sizes that exceed 0.1. Furthermore, the blue solid line illustrates the fractional size $S$ of the giant cluster over 1000 hypergraph realizations.
}
    \label{fig:cascade_window}
\end{figure*}

The cascade condition in Eq.~\eqref{eq:threshold_condition} can be understood as follows: If $G_{k1}^{\prime}(1)G_{m1}^{\prime}(1) < 1$, all vulnerable clusters within the system remain small and incapable of initiating a global cascade. Conversely, when $G_{k1}^{\prime}(1)G_{m1}^{\prime}(1) > 1$, a single, infinite-sized \emph{giant vulnerable cluster} emerges as $N \to \infty$, possessing the potential to trigger a global cascade, albeit with a finite probability. As an illustrative example, Fig.~\ref{fig:cascade_window}(a) depicts a cascade region, delineated by a red line, which is determined by Eq.~\eqref{eq:threshold_condition}. This line marks the boundary between two distinct phases, with a phase transition occurring upon its traversal. In particular, global cascades can only occur when both $\phi_k$ and $\langle k \rangle$ assume lower values. It's noteworthy that in the case of a single seed node, the cascade condition specified in Eq.~\eqref{eq:threshold_condition} indicates an equal influence or impact of two factors, hyperdegree and cardinality, on the cascade process. It means that they play equally important roles and symmetrically contribute to the dynamics of the cascade.


As proposed by  \cite{watts2002simple}, within the cascade region, the fractional size $S_v$ of the giant vulnerable cluster alone tends to underestimate the probability of a global cascade. This underestimation arises because even stable nodes can trigger global cascades if they are adjacent to the vulnerable cluster (see Fig.~\ref{fig:volunerale_cluster} in Appendix ~\ref{sec:app_A}). Therefore, the true boundary of the cascade region should be determined by evaluating the fractional size $S_e$ of the \emph{extended giant vulnerable cluster} (see the boundary of the black area in Fig.~\ref{fig:cascade_window}(a), or the red dashed line in Fig.~\ref{fig:cascade_window}(b)). In addition, the fractional size $S_e$ of the extended giant vulnerable cluster is calculated by $S_e = S_c + S_v$, where $S_c$ is the fractional size of \emph{complementary giant vulnerable cluster}. The analytical solutions for $S_v$ and $S_c$ can be referred as to Eqs.~\ref{eq:S_v} and \ref{eq:S_c} in Appendix ~\ref{sec:app_A}. 

In Fig.~\ref{fig:cascade_window}(b), three curves depict the fractional sizes of the giant vulnerable cluster, $S_v$, the extended giant vulnerable cluster, $S_e$, and the cascade size, $S$, as the average hyperdegree $\langle k \rangle$ varies. The value of $S_v$ can be derived from Eq.~\eqref{eq:S_e} in Appendix \ref{sec:app_A} (yellow dotted line). Additionally, the value of $S_e$ is a good approximation of the frequency of global cascades (the red circles). It is worth noting that the average cascade size (the blue triangles) is not governed by the value of $S_v$ or $S_e$, but by the average size $S$ of the giant cluster (the blue line). Initially, for lower values of $\langle k \rangle$, the fractional sizes of these clusters overlap. However, around $\langle k \rangle = 3$, a noticeable gap emerges between the giant vulnerable cluster and the other two clusters. Subsequently, around $\langle k \rangle = 7$, a similar gap arises between the extended giant vulnerable cluster and the cascade size. Finally, as $\langle k \rangle$ exceeds approximately 12, the values of $S_v$ and $S_e$, and the cascade size $S$, decrease to 0. This phenomenon occurs because the hypergraph's extensive connectivity renders the whole giant cluster invulnerable. Note that the size of the giant cluster will not be 0 (blue solid line) since it only captures the network structure, not the information cascade dynamics.

\subsection{Any Seed Size}
We next investigate the general case where the initial fractional seed size $r_0\in [0,1] $ can take any value. 

We focus here on the average fraction \( r \) of active nodes in the steady state, which represents the ensemble average over realizations of hypergraphs. It is important to note that if the initial fractional seed size $r_0 \in (0,1]$ cannot be neglected, the value of \( r \) differs from the average cascade size \( S \) in the case of a single seed node, i.e., $r_0 = 1/N \to 0$. This difference arises because the initial seed nodes can be positioned in separate clusters, not exclusively within the giant cluster. Thus the seed nodes can activate a fraction of nodes in these smaller clusters, contributing to the overall activation size.

The average fraction \( r \) of active nodes in the steady state can be given by
\begin{equation}
    r = r_0 + (1-r_0) \sum_{k=1}^\infty p_k \sum_{i=0}^k \binom{k}{i} u_{\infty}^i (1-u_{\infty})^{k-i}F_k(i),
    \label{eq:r}
\end{equation}
where 
\begin{equation}
F_k(i)=\left\{
     \begin{array}{lr}
    0, & \text{ if } i/k < \phi_k, \\
     1, & \text{ if } i/k  \geq  \phi_k.
     \end{array}
\right.
\end{equation}

Here $u_\infty$ is the probability that a random hyperedge reached by a random associated node is active, which is the fixed point of the following recursive equation
\begin{equation}
    u_{n+2} = g\Big(r_0+(1-r_0)f(u_n)\Big).
\label{eq:recusive}
\end{equation}

The nonlinear function $g$ is defined as 
\begin{equation}
    g(w)=\sum\limits_{m=1}^\infty \dfrac{m q_m}{\langle m \rangle} \sum\limits_{i=0}^{ m-1} \binom{m-1}{i} w^{i}(1-w)^{m-1-i} F_m(i),
    \label{eq:g(w_n)}
\end{equation}
where 
\begin{equation}
F_m(i)=\left\{
     \begin{array}{lr}
          0, & \text{ if } i/m < \phi_m, \\
     1, & \text{ if } i/m  \geq   \phi_m.
     \end{array}
\right.
\end{equation}

Here $w$ represents the probability that a random node reached by a random hyperedge is active. Moreover, we have
\begin{equation}
    w_{n+1} = r_0+(1-r_0)f(u_n),
    \label{eq:w_n+}
\end{equation}
with $w_0=r_0$ and the nonlinear function $f$ defined as
\begin{equation}
    f(u)=\sum\limits_{k=1}^\infty \dfrac{k p_k}{\langle k \rangle} \sum\limits_{i=0}^{k-1} \binom{k-1}{i} u^{i} (1-u)^{k-1-i}F_k(i),
    \label{eq:f(u_n)}
\end{equation}
with $u_1=g(w_0)$. The logic and details of the derivation of the above equations are illustrated in Fig.~\ref{fig:illu_any_seed} in Appendix ~\ref{sec:active_nodes} and described in further detail in Appendix ~\ref{sec:active_nodes}.

If some initial seed nodes are randomly positioned in the giant cluster and subsequently induce a significant number of nodes to become active, we say a global cascade occurs. Writing $f(u)$ in Eq.~\eqref{eq:f(u_n)} as $\sum_{l_1=0}^\infty C_{l_1} u^{l_1}$ and $g(w)$ in Eq.~\eqref{eq:g(w_n)} as $\sum_{l_2=0}^\infty B_{l_2} w^{l_2}$, and linearizing Eq.~\eqref{eq:recusive}, we can obtain the following cascade condition 
\begin{equation}
    \sum_{k=1} ^\infty \frac{k (k-1)p_k}{\langle k \rangle}\Big[F_k(1) - F_k(0)\Big] \sum_{l_2=1} l_2 B_{l_2} r_0^{l_2-1} > \frac{1}{1-r_0}.
    \label{eq:double_condition}
\end{equation}

\begin{figure*}[htbp]
    \centering
    \includegraphics[width=\linewidth]{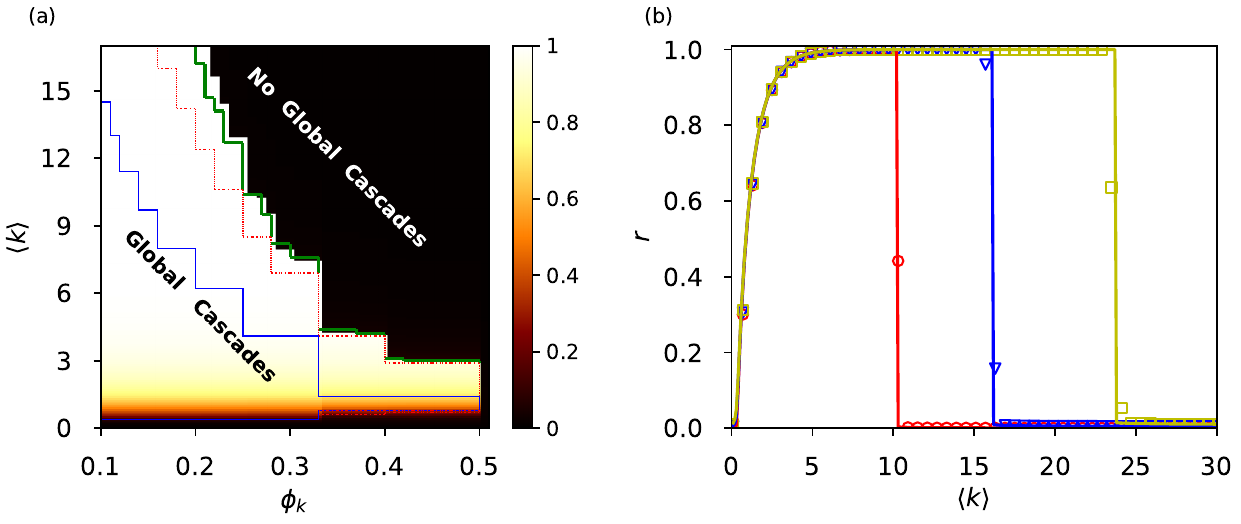}
    \caption{The average fraction $r$ of active nodes in random hypergraphs. (a) Color-coded values of $r$ in the simulations on the $(\phi_k, \langle k \rangle)$ plane with $r_0 = 10^{-2}$, $\phi_m = 0.1$, $\langle m \rangle = 3$, and $N = 10^5$; (b) Values of $r$ at $\phi_k = 0.18$ and $\phi_m = 0.1$ from Eq.~\eqref{eq:r} (lines) and numerical simulations (points), averaged over 100 realizations with $N = 10^5$ for different fractional seed sizes $r_0 = 10^{-3}$, $5 \times 10^{-3}$, $10^{-2}$.}
        \label{fig:cascade_window_r}
\end{figure*}

\begin{figure*}
    \centering
    \includegraphics[width=\linewidth]{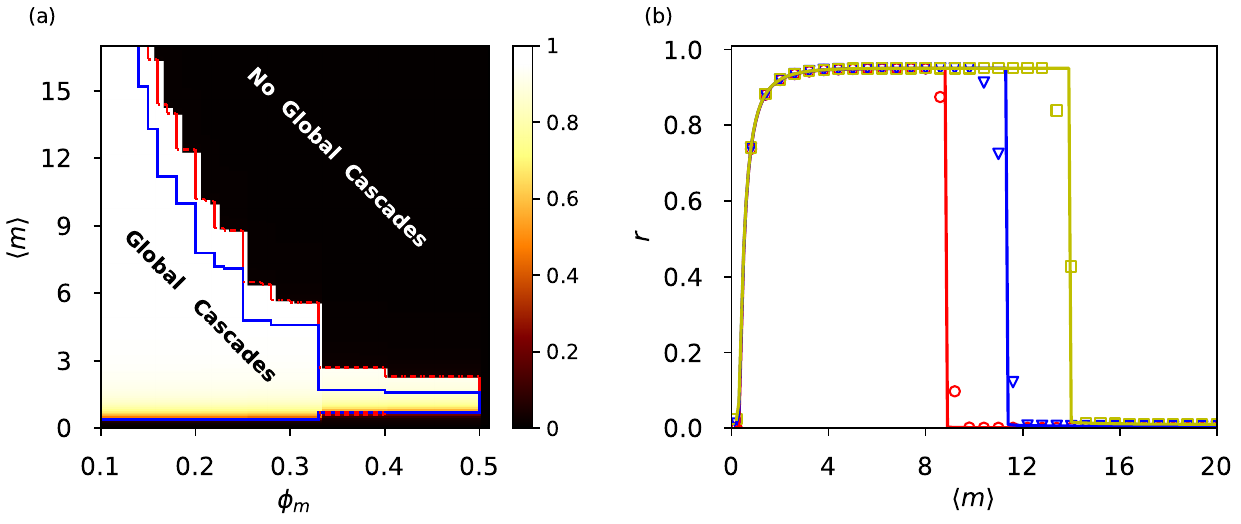}
    \caption{The average fraction $r$ of active nodes in random hypergraphs. (a) Color-coded values of $r$ in the simulations on the $(\phi_m, \langle m \rangle)$ plane with $r_0=10^{-2}$, $\phi_k=0.1$, $\langle k \rangle =3$,  and $N=10^5$; (b) Values of $r$ at $\phi_m=0.18$ and $\phi_k=0.1$ from Eq.~\eqref{eq:r} (lines) and numerical simulations (points), averaged over 100 realizations with $N=10^5$ for different fractional seed sizes $r_0 = 10^{-3}$, $5 \times 10^{-3}$, $10^{-2}$.}
    \label{fig:cascade_window_r_phim}
\end{figure*}
The details of the derivation of the above cascade condition can be found in Appendix~\ref{sec:cascade_condition_de}. On the $(\phi_k, \langle k \rangle)$ plane depicted in Fig.~\ref{fig:cascade_window_r}(a), this condition holds true within the region delineated by the blue solid line. It's evident that the condition outlined in Eq.~\eqref{eq:double_condition} inadequately represents the actual boundary of global cascades. As argued in \cite{gleeson2007seed}, an improved condition needs to be sought by extending the series for Eq.~\eqref{eq:h(r_0,u)} in Appendix~\ref{sec:cascade_condition_de} to higher orders.

On simple networks \cite{gleeson2007seed}, the series is extended to a square polynomial function $b u^2 + c u + d = 0$, which can accurately capture the true boundary of the global cascade. They thus extend the cascade boundary to include regions where either $c > 0$ or $c^2 - 4bd < 0$. However, for hypergraphs in our model, the improved cascade condition involves a cubic polynomial function. Extending the series to a square polynomial does not accurately determine the boundary of the global cascade (see the red dotted line), but extending the series to a cubic polynomial function of the form $a u^3 + b u^2 + c u + d = 0$ does (see the green dashed line). Subsequently, imposing the constraint that the cubic polynomial function has no solution yields the boundary, represented by the green dashed line.

Furthermore, in the limit as $r_0 \to 0$, with $F_k(0) = F_m(0) = 0$, the condition specified in Eq.~\eqref{eq:threshold_condition} emerges. It is crucial to recognize that, unlike the case involving a single seed node $r_0 = 1/N \to 0$, when $r_0$ attains a significant level, hyperdegree $k$ and cardinality $m$ exert unequal degrees of influence on the cascade condition, as delineated in Eq.~\eqref{eq:double_condition}.

In Figs.~\ref{fig:cascade_window_r}(a) and \ref{fig:cascade_window_r_phim}(a), when we set $\phi_m = \phi_k = 0.1$, the cascade boundaries delineated by the blue solid lines differ markedly. Specifically, in Fig.~\ref{fig:cascade_window_r_phim}(a), the boundary expands compared to Fig.~\ref{fig:cascade_window_r}(a). However, the actual upper boundary of the global cascade, represented by the green dashed line in Fig.~\ref{fig:cascade_window_r_phim}(a), is higher than that delineated by the red dotted line in Fig.~\ref{fig:cascade_window_r}(a). 

This observation is further supported by the positions of critical points for the first-order phase transitions in Figs.~\ref{fig:cascade_window_r}(b) and \ref{fig:cascade_window_r_phim}(b). Specifically, we set symmetric parameter values: $\phi_k = 0.18$ and $\phi_m = 0.1$ for one set of experiments, and $\phi_k = 0.1$ and $\phi_m = 0.18$ for another. We find that in Fig.~\ref{fig:cascade_window_r}(b), the critical points for the average hyperdegree $\langle k \rangle$ are approximately 10, 16, and 24, while in Fig.~\ref{fig:cascade_window_r_phim}(b), they are about 9, 12, and 13 for the same initial seed size. This indicates that under symmetric threshold settings for $\phi_k$ and $\phi_m$, average hyperdegree $\langle k \rangle$ triggers the global cascade more frequently than the effect of average cardinality $\langle m \rangle$.

\section{Conclusion}
This study introduces a comprehensive framework that positions the research on information cascades within the context of higher-order interactions, leveraging a double-threshold hypergraph model. We elucidate the conditions for information dissimination among individuals (nodes) through each communication channel (hyperedge): individuals become aware of the information once the number of information adopters surpasses a threshold $\phi_m$; however, actual adoption occurs only when the cumulative influence across all communication channels exceeds another threshold $\phi_k$.  
  
Employing percolation methods and mean-field approximation, we derive the cascade conditions for both a single seed node and any seed size. Our analysis reveals the asymmetric impact of the connectivity pattern of the random hypergraph, characterized by the distributions of hyperdegree $k$ and cardinality $m$, on the global cascade boundary. This asymmetry manifests in the observed differences in the boundaries of the global cascade within the $(\phi_m, \langle m \rangle)$ and $(\phi_k, \langle k \rangle)$ planes.  Notably, as the initial seed size $r_0$ approaches 0, this asymmetric effect diminishes.

These findings significantly contribute to our comprehension of information diffusion processes within higher-order complex systems. Nonetheless, our analysis leans on a theoretical framework, which may not fully capture the intricacies of real-world networks. Integrating empirical data and conducting validation studies would bolster the applicability of our findings. Additionally, while our model provides valuable insights, its simplifications may obscure the complexities inherent in information cascades. Thus, exploring more sophisticated models that account for heterogeneous node attributes and dynamic network structures is warranted. Furthermore, our focus on homogeneous networks overlooks the dynamics of heterogeneous networks, representing a compelling avenue for future research exploration.

\section{Acknowledgments}
    This work is supported by the National Natural Science Foundation of China (Grant No. 61773148 and 52374013) and the Entrepreneurship and Innovation Project of High Level Returned Overseas Scholar in Hangzhou.

\appendix

\section{Exdended Vulnerable Cluster \label{sec:app_A}}

\begin{figure}[htbp]
    \centering
    \includegraphics[width=\linewidth]{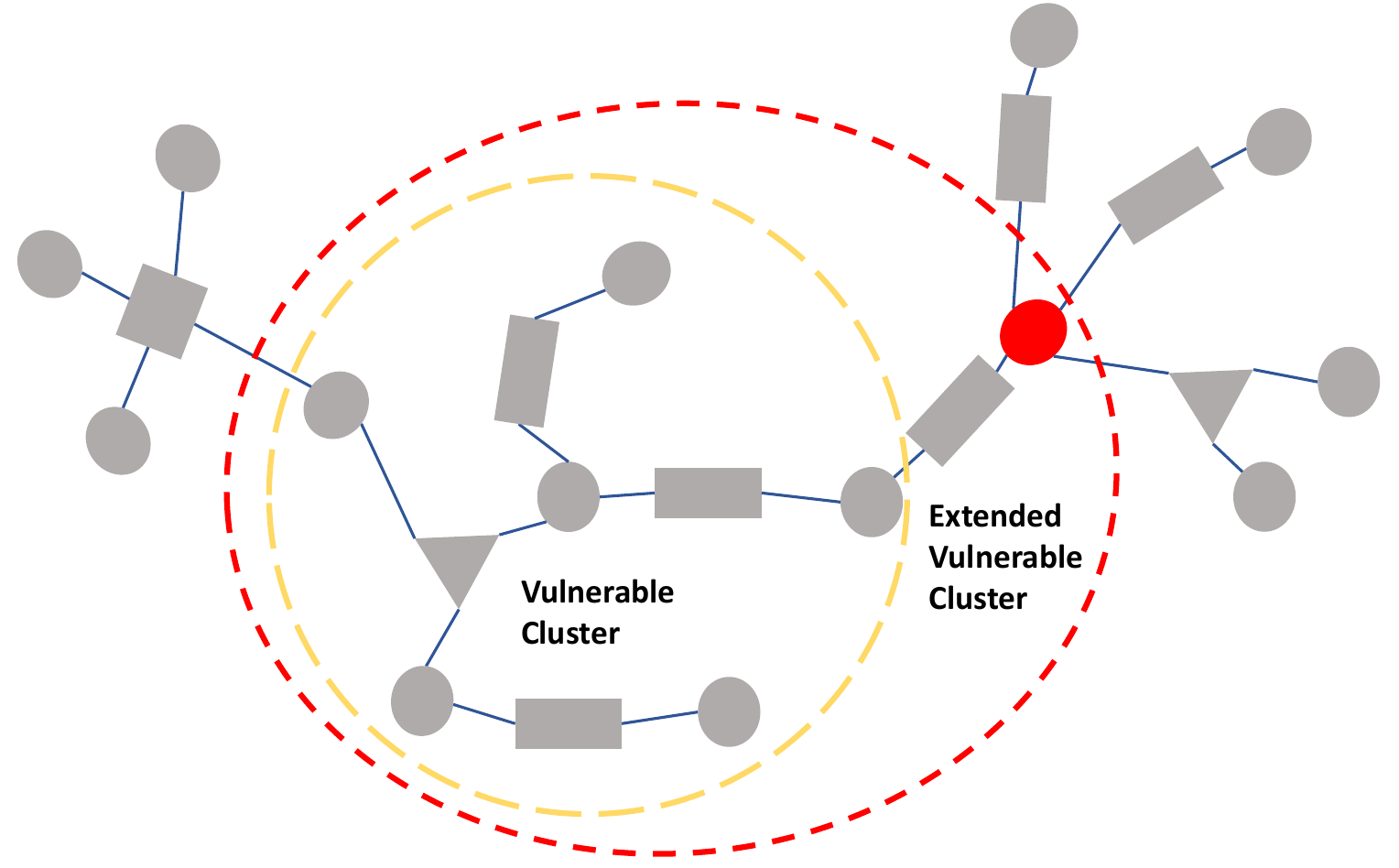}
    \caption{A schematic illustration of vulnerable cluster with $\phi_k = 0.3$ and $\phi_m = 0.3$. The nodes inside the yellow circle constitute the vulnerable cluster, while the red circle represents the extended vulnerable cluster including the red node.}
    \label{fig:volunerale_cluster}
\end{figure}
We can calculate the fractional size $S_v$ of the giant vulnerable cluster within the cascade window using the following expression
\begin{equation}
S_v = 1-H_0(1).
\label{eq:S_v}
\end{equation}

Furthermore, the fractional size $S_c$ of the complementary giant vulnerable cluster can be computed as follows
\begin{equation}
S_c = \sum_k p_{k} (1-\rho_{k})\Big[1-(1-R)^{k}\Big],
\label{eq:S_c}
\end{equation}
where $R$ is the probability that a node, through a vulnerable hyperedge, can reach a vulnerable cluster.


Additionally, the probability $R$ can be determined by
\begin{equation}
\begin{split}
R = & \sum_{m=1} \frac{ \beta_m q_m m }{\langle m \rangle} \sum_{n=0}\binom{m-1}{n}G_{k1}^{n}(1)\Big(1-G_{k1}(1)\Big)^{m-n-1} \\
& \times \Big[1-\frac{G_{k1}^{n}(1-R)}{G_{k1}^n(1)}\Big].
\end{split}
\label{eq:R}
\end{equation}

By integrating Eqs.~\eqref{eq:S_v} and \eqref{eq:S_c}, we can determine the fractional size $S_e$ of extended giant vulnerable cluster by
\begin{equation}
    S_e = S_c + S_v. 
    \label{eq:S_e}
\end{equation}

\section{Comparison with Simple Networks}
In the case of a single node (i.e., $r_0 = 1/N \to 0$), note that when $m = 2$ (i.e., simple networks) and $\phi_m \leq 0.5$, our model reduces to the model of \cite{watts2002simple}. Thus, we can obtain $G_{m1}(x) = x$ in Eq.~\eqref{eq:G_m1}. Additionally, Eqs.~\eqref{eq:H_1(x)} and \eqref{eq:h_0(x)} can be reduced to the equations in the work of \cite{watts2002simple} as
\begin{equation}
\begin{cases}
    H_1(x) = 1 - G_1(1) + x G_1(H_1(x)),\\
    H_0(x) = 1 - G_0(1) + x G_0(H_1(x)).
\end{cases}
\end{equation}

Further, we get the cascade condition in the simple network as
\begin{equation}
    \frac{\sum_{k} k(k-1) \rho_{k} p_{k}}{\langle k \rangle} = 1.
\end{equation}

Moreover, in the case of any seed size (i.e., $r_0 \in [0,1]$), note that when $m = 2$ (i.e., simple networks) and $\phi_m \leq 0.5$, our model reduces to the model of \cite{gleeson2007seed}. Thus, Eq.~\eqref{eq:g(w_n)} reduces to $g(w) = w$, and Eq.~\eqref{eq:recusive} becomes
\begin{equation}
    u_{n+1} = r_0 + (1 - r_0)f(u_n).
\end{equation}

Therefore the cascade condition will be 
\begin{equation}
    \sum_{k=1} ^\infty \frac{k (k-1)p_k}{\langle k \rangle}\Big[F_k(1) - F_k(0)\Big] > \frac{1}{1-r_0}.
\end{equation}

\begin{figure}[htbp]
    \centering
    \includegraphics[width=\linewidth]{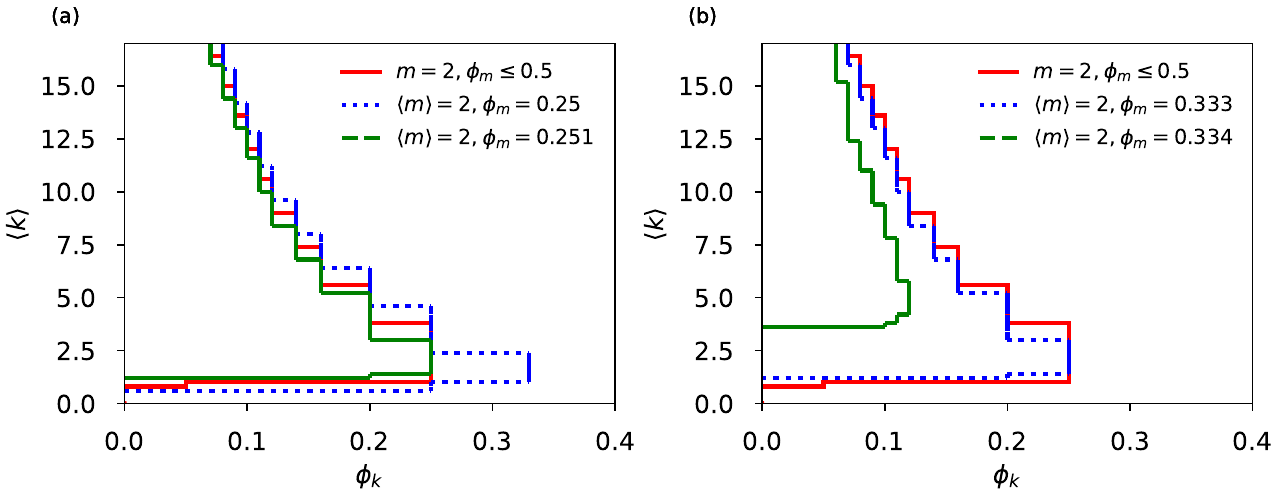}
    \caption{(a) The cascade regions with fixed cardinality and average cardinality for different threshold values of $\phi_m$ in the case of a single seed node. The red solid line represents the case with fixed cardinality $m = 2$ and $\phi_m \leq 0.5$ (i.e., the model of \cite{watts2002simple}), while the blue and green lines represent the cases with a Poisson cardinality distribution with average cardinality $\langle m \rangle = 2$ and $\phi_m = 0.25$ and $\phi_m = 0.251$, respectively; (b) The red solid line represents the case with fixed cardinality $m = 2$ and $\phi_m \leq 0.5$ (i.e., the model of \cite{watts2002simple}), while the blue and green lines represent the cases with a Poisson cardinality distribution with average cardinality $\langle m \rangle = 2$ and $\phi_m = 0.333$ and $\phi_m = 0.334$, respectively.
}
    \label{fig:fixm}
\end{figure}

The results shown in Fig.~\ref{fig:fixm} indicate that the threshold value $\phi_m$ has a significant impact on the cascade region for hypergraphs with a Poisson cardinality distribution. Specifically, in Fig.~\ref{fig:fixm}(a), when the average cardinality is $\langle m \rangle = 2$, $\phi_m \leq 0.5$ (i.e., the model of \cite{watts2002simple}), $\phi_m = 0.25$ includes the hyperedges of cardinality $m = 4$ as vulnerable, resulting in the largest cascade region (blue line). In contrast, $\phi_m = 0.251$ does not include these hyperedges, leading to the smallest cascade region (green line). Similarly, in Fig.~\ref{fig:fixm}(b), for $\langle m \rangle = 2$, $\phi_m = 0.333$ includes hyperedges of cardinality $m = 3$ as vulnerable, producing the largest cascade region (blue line), whereas $\phi_m = 0.334$ excludes these hyperedges, resulting in the smallest cascade region (green line). Compared to the fixed cardinality of $m = 2$ (red line), only the case with $\phi_m = 0.25$ significantly expand the cascade region, while $\phi_m = 0.251$, $\phi_m = 0.333$, and $\phi_m = 0.334$ reduce them.

\begin{figure}[htbp]
    \centering
    \includegraphics[width=\linewidth]{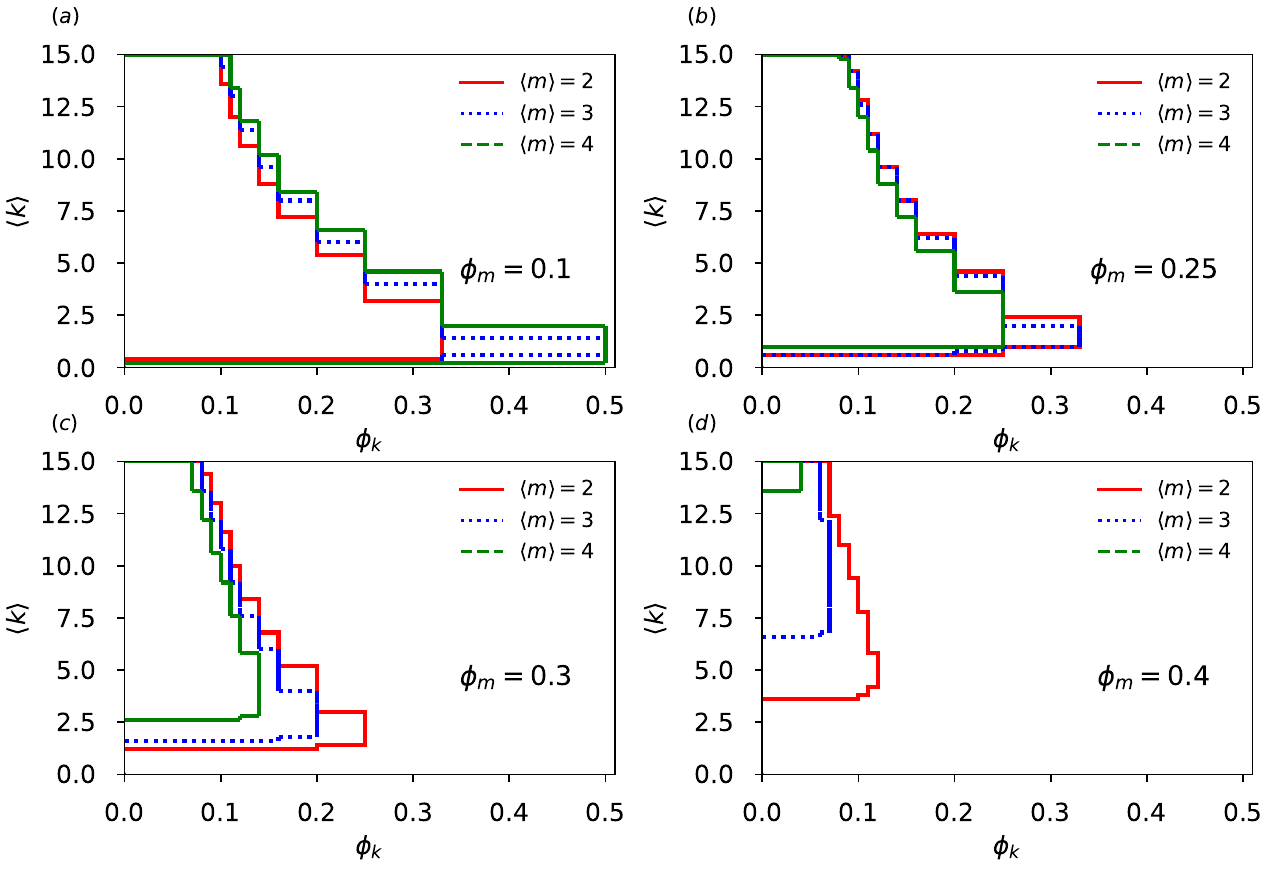}
    \caption{The cascade regions are depicted for varying average hyperedge cardinality $\langle m \rangle$ in the case of a single seed node. Panels (a-d) correspond to distinct hyperedge thresholds $\phi_m$, with values of 0.1, 0.25, 0.3, and 0.4, respectively. These panels offer insight into the shifting and narrowing of the cascade region as $\phi_m$ increases.
}
    \label{fig:vary_m}
\end{figure}

In Fig.~\ref{fig:vary_m}, the cascade region's sensitivity to changes in the average hyperedge cardinality $\langle m \rangle$ becomes evident. At a lower hyperedge activation threshold (e.g., $\phi_m = 0.1$), increasing $\langle m \rangle$ noticeably expands the cascade region. This expansion is due to the heightened connectivity of the hypergraph, which engages more nodes in the cascade process, making most hyperedges vulnerable. Conversely, at a higher threshold (e.g., $\phi_m = 0.4$), the trend reverses. Specifically, higher values of $\langle m \rangle$ result in a more restricted cascade region. This occurs because a higher threshold (e.g., $\phi_m = 0.4$) makes more hyperedges stable, especially with larger $\langle m \rangle$, posing a significant challenge for initiating a global cascade.

\section{Heterogeneous Thresholds}
\begin{figure}[htbp]
    \centering
    \includegraphics[width=\linewidth]{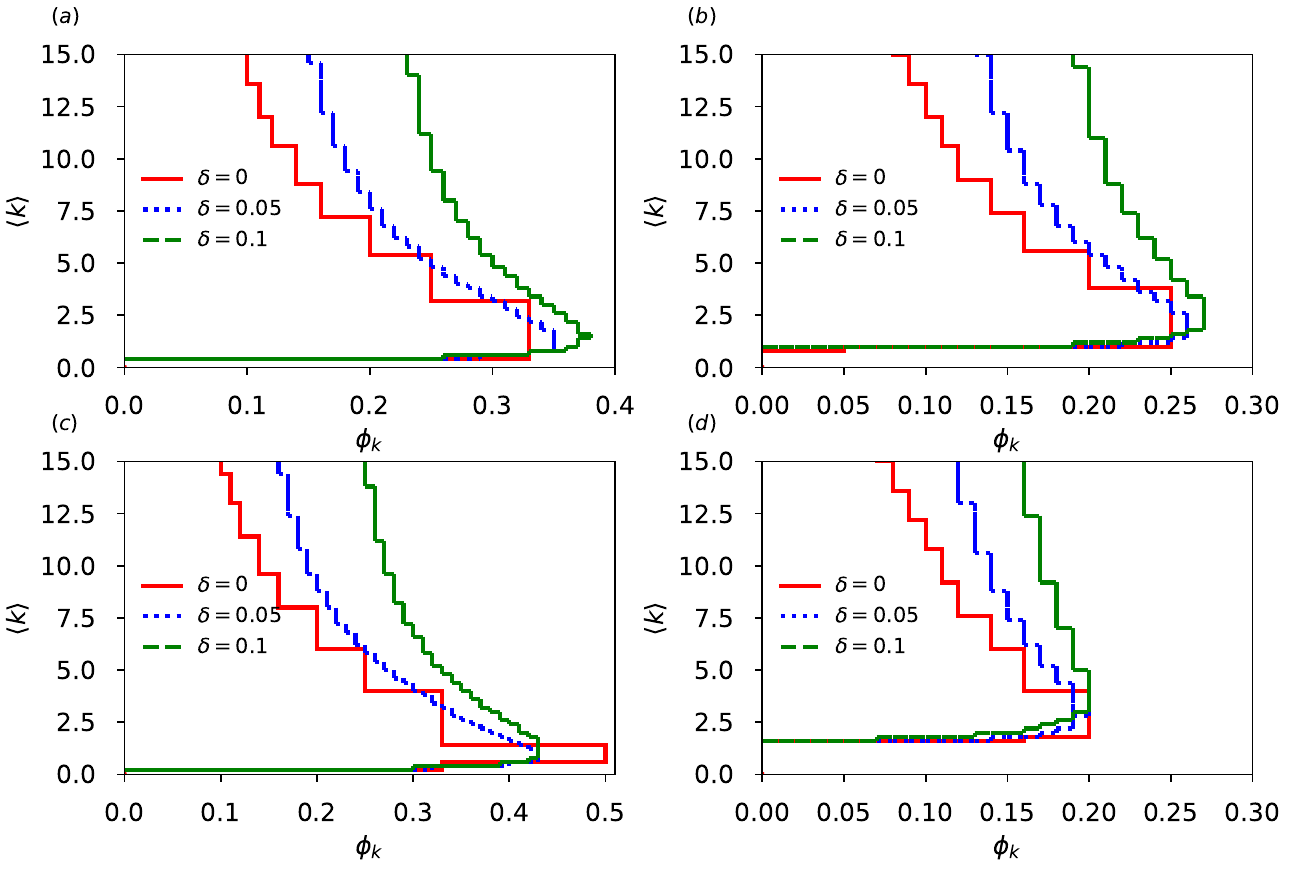}
    \caption{Analytically derived cascade regions by Eq.~\eqref{eq:threshold_condition}with single seed node in random hypergraphs for homogeneous $\delta=0$ and heterogeneous thresholds, where $\phi_k$ follows a normal distribution with standard deviations $\delta=0.05$ and $\delta=0.1$. The cases considered are as follows: (a) $\langle m \rangle=2$, $\phi_m=0.1$; (b) $m=2$, $\phi_m \leq 0.5$; (c) $\langle m \rangle=3$, $\phi_m=0.1$; (d) $\langle m \rangle=3$, $\phi_m=0.251$. }
    \label{fig:uneven_threshold}
\end{figure}
     
In Fig.~\ref{fig:uneven_threshold}, the cascade region is depicted for a homogeneous threshold $\phi_m$ (red line) alongside two other cascade regions (blue and green lines) representing threshold distributions that follow a normal distribution with the same average $\phi_m$ and standard deviations $\delta = 0.05$ and $0.1$, respectively. Increased heterogeneity in thresholds leads to decreased system stability, resulting in cascades occurring over a wider range of both $\phi_k$ and $\langle k \rangle$ (see Fig.~\ref{fig:uneven_threshold}(a-d)). Similar to the homogeneous threshold cases, under a lower value of $\phi_m = 0.1$, a higher value of $\langle m \rangle = 3$ can trigger a larger cascade region in inhomogeneous threshold cases as well (see the blue and green lines in Fig.~\ref{fig:uneven_threshold}(a,c)). Additionally, for inhomogeneous threshold cases, a larger value of $\phi_m = 0.251$ (excluding hyperedges with $m = 4$ as vulnerable) can greatly increase system stability, as illustrated by the shrunken cascade region (see the blue and green lines in Fig.~\ref{fig:uneven_threshold}(c,d)).

\section{Derivation of Eq.~\eqref{eq:r} \label{sec:active_nodes}}

\begin{figure}[h]
    \centering
    \includegraphics[width=0.9\linewidth]{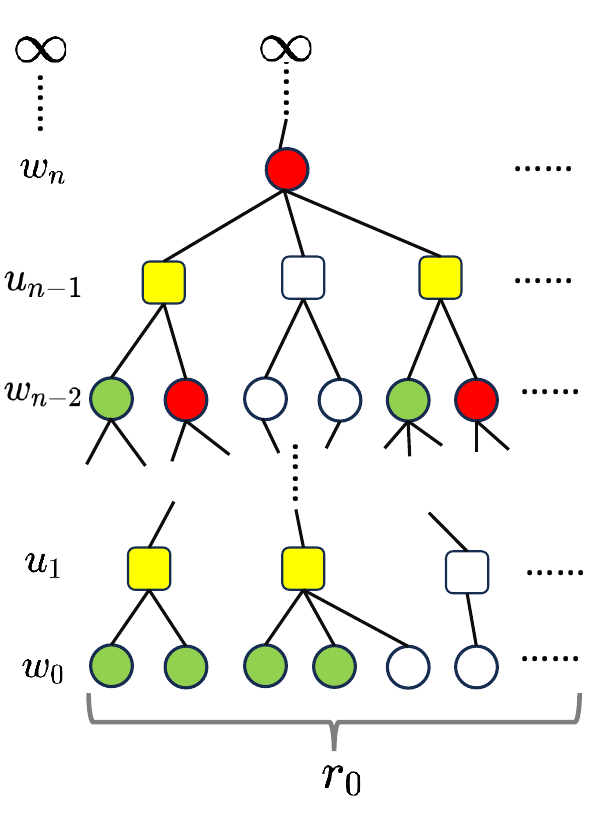}
        \caption{A schematic illustration of the derivation of the function in Eq.~\eqref{eq:r}. The random hypergraph can be treated as a tree structure. Specifically, the circles represent nodes, with green denoting seed nodes, red indicating activated nodes during the cascade process, and white representing inactive nodes. Additionally, squares denote hyperedges, with yellow indicating activation during the cascade process and white remaining inactive. }
    \label{fig:illu_any_seed}
\end{figure}

The logic of derivation of Eq.~\eqref{eq:r} can be illustrated in Fig.~\ref{fig:illu_any_seed}. Here, the random hypergraph can be treated as a tree structure. The topmost level comprises a single node with hyperdegree $k$, connected to its $k$ hyperedges at the next level. Each hyperedge, in turn, contains $m-1$ neighboring nodes at the subsequent lower level. The cardinality distribution of the hyperedges is given by $\frac{mq_m}{\langle m \rangle}$. Furthermore, each node in a hyperedge connects to $k-1$ hyperedges at the next lower level, with the hyperdegree distribution given by $\frac{k p_k}{\langle k \rangle}$. To determine the final fraction of active nodes, we label the tree levels from the bottom ($n=0$) to the top ($n=\infty$). Let $u_n$ denote the conditional probability that a hyperedge at level $n$ is active, given that the upper-level nodes it contains are inactive (at level $n+1$). A node will be activated in two cases: it is a seed node or the number $i$ of active hyperedges in Eq.~\eqref{eq:f(u_n)} exceeds the threshold $\phi_k$, i.e., $F_k(i) = 1$. Thus, we obtain Eq.~\eqref{eq:w_n+}. Similarly, the derivation for level $n+2$ can be given by Eq.~\eqref{eq:g(w_n)}. Finally, the probability that the single node at the tree's top is active arises from two cases: it is already active as a seed node, or it becomes active if a sufficient number of its $k$ hyperedges are active. This leads to Eq.~\eqref{eq:r}.

\section{Cascade Condition \label{sec:cascade_condition_de}}
As outlined in  \cite{gleeson2007seed}, we can express $f(u)$ in Eq.~\eqref{eq:f(u_n)} as $\sum_{l_1=0}^\infty C_{l_1} u^{l_1}$ with coefficients
\begin{equation}
    C_{l_1} = \sum_{k=l_1+1}^\infty \sum_{n=0}^{l_1} \binom{k-1}{l_1} \binom{l_1}{n} (-1)^{l_1+n} \frac{k p_k}{\langle k \rangle} F_k(n).
\end{equation}

Similarly, $g(w)$ in Eq.~\eqref{eq:g(w_n)} can be represented as $\sum_{l_2=0}^\infty B_{l_2} w^{l_2}$ with coefficients
\begin{equation}
    B_{l_2} = \sum_{m=l_2+1}^\infty \sum_{n=0}^{l_2} \binom{m-1}{l_2} \binom{l_2}{n} (-1)^{l_2+n} \frac{m q_m}{\langle m \rangle} F_m(n).
\end{equation}

Subsequently, Eq.~\eqref{eq:recusive} can be reformulated as
\begin{equation}
    \sum_{l_2=0}^\infty B_{l_2} \Big(r_0 + (1-r_0) \sum_{l_1=0}^\infty C_{l_1} u^{l_1} \Big)^{l_2} - u = 0,
    \label{eq:simple_u}
\end{equation}
with the function defined as
\begin{equation}
    h(r_0,u) = \sum_{l_2=0}^\infty B_{l_2} \Big(r_0 + (1-r_0) \sum_{l_1=0}^\infty C_{l_1} u^{l_1} \Big)^{l_2} - u.
    \label{eq:h(r_0,u)}
\end{equation}

Neglecting the influence of $O(u^2)$ of Eq.~\eqref{eq:h(r_0,u)}, since $u_n$ increases with $n$ at least initially, the condition $h'(r_0,0) - 1 = (1-r_0) C_1 \sum_{l_2=1} l_2 B_{l_2} r_0^{l_2-1} - 1 > 0$ must be satisfied. 

Thus we can obtain the following condition 
\begin{equation}
    \sum_{k=1} ^\infty \frac{k (k-1)p_k}{\langle k \rangle}\Big[F_k(1) - F_k(0)\Big] \sum_{l_2=1} l_2 B_{l_2} r_0^{l_2-1} > \frac{1}{1-r_0}.
\end{equation}



\section{Critical Seed Size}

With Eq.~\eqref{eq:h(r_0,u)} in Appendix ~\ref{sec:cascade_condition_de}, the critical solutions of $r_{0,c}$ and $u_c$ satisfy the following conditions
\begin{equation}
    \begin{cases}
        \frac{\partial h}{\partial u} (r_{0,c}, u_c) = 0, \\
        h(r_{0,c}, u_c) = 0.
    \end{cases}
    \label{eq:phase_transition_condition}
\end{equation}

\begin{figure}
    \centering    \includegraphics[width=\linewidth]{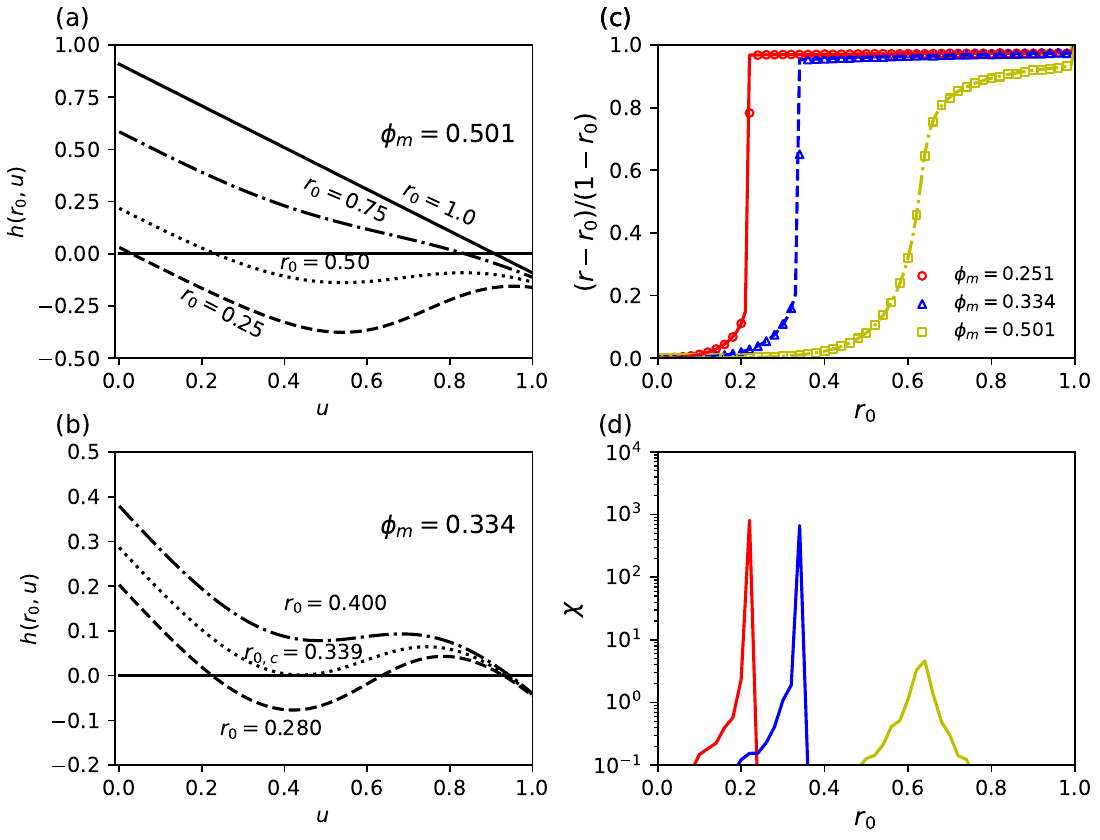}
    \caption{(a-b) The graphical solutions for Eq.~\eqref{eq:h(r_0,u)} are presented across varied values of $\phi_m$ within a random hypergraph, while $\langle m \rangle=4$, $\phi_k=0.501$; (c-d) The relationship between the complementary fractional cascade size $(r-r_0)/(1-r_0)$ and the system susceptibility $\chi$ is depicted for various initial fractional seed sizes $r_0$, across different $\phi_m$ values in random hyperdegrees with $\phi_k=0.501$ and $\langle m \rangle=4$. The lines denote analytical results, while the points represent simulation outcomes.}
    \label{fig:r_r0}
\end{figure}

The function curves representing $h(r_0,u)$ for various parameter configurations are depicted in Fig.~\ref{fig:r_r0}(a,b). For a higher value of $\phi_m$ (e.g., $\phi_m=0.501$), a single solution is observed irrespective of the fractional seed size $r_0$ (see Fig.~\ref{fig:r_r0}(a)). However, with a lower $\phi_m$ (e.g., $\phi_m=0.334$), nontrivial solutions for $h(r_0,u)$ emerge as $r_0$ increases, causing the function curve to become tangent to the horizontal axis (see Fig.~\ref{fig:r_r0}(b)).

Consequently, we identify a first-order percolation transition point $\rho_{0,c}$, where $u$ undergoes an abrupt transition from a lower value $u_c$ to a higher value $u_{c2}$. This results in a discontinuous increase in the complementary fractional cascade size $(r-r_0)/(1-r_0)$, as depicted in Fig.~\ref{fig:r_r0}(c). Additionally, the corresponding susceptibility $\chi$ reaches a peak value, as shown in Fig.~\ref{fig:r_r0}(d).

\bibliographystyle{unsrt}
\bibliography{main.bib}
\end{document}